\documentclass[aps,pre,floats]{revtex4}
\usepackage{graphicx}
\usepackage{amssymb}
\begin{document}
\title{Front Propagation up a Reaction Rate Gradient}

\author{Elisheva Cohen} 
\affiliation{Dept. of Physics, Bar-Ilan  University, Ramat-Gan, IL52900 Israel}
\author{David A. Kessler}
\affiliation{Dept. of Physics, Bar-Ilan University, Ramat-Gan, IL52900 Israel}
\author{Herbert Levine} 
\affiliation{Center for Theoretical Biological
  Physics, University of California, San Diego, 9500 Gilman Drive, La
  Jolla, CA 92093-0319} 
\date{\today}
\begin{abstract}
  We expand on a previous study of fronts in finite particle
  number reaction-diffusion systems in the presence of a
  reaction rate gradient in the direction of the front motion. 
  We study the system via reaction-diffusion equations, 
  using the expedient of a cutoff in the reaction rate
  below some critical density to capture the essential role of fluctuations in the
  system. For large density, the velocity is large, which allows for an
  approximate analytic treatment.
  We derive an analytic approximation for the front velocity dependence on
  bulk particle density, showing that the velocity indeed diverges in the
  infinite density limit.  The form in which diffusion is implemented, namely nearest-neighbor
  hopping on a lattice,  is seen to have
  an essential impact on the nature of the divergence.
\end{abstract}
\maketitle

\section{Introduction}

The propagation of fronts connecting different macroscopic states is a common
occurrence in many non-equilibrium systems~\cite{vansaarloos-review}. Familiar 
examples range from solidification~\cite{kkl} to chemical reaction dynamics 
such as flames~\cite{kpp} and to the spatial spread of
infections~\cite{blumen} through a susceptible population. Previous work by 
many authors has shown that a useful way to classify such fronts is via the 
stability properties of the state being invaded. In fact, surprising 
differences, with regard to the selection of the speed of 
propagation~\cite{ben-jacob}, the rate of approach to that 
speed~\cite{bd,vansaarloos,kns}, the sensitivity to finite-particle number 
fluctuations~\cite{bd,pechenik}, and the stability to 2d 
undulations~\cite{nature}, exist between fronts that propagate into metastable 
versus linearly unstable states.

In a previous work~\cite{kess}, we addressed the question of the
dynamics a new type of front, that which exists in a  system containing a 
reaction-rate gradient in the
direction of front motion~\cite{shapiro}.  Our staring point was a simple 
infection model  $A+B \rightarrow 2A$ on a
1d lattice (with spacing $a$) with equal $A$ and $B$ hopping 
rates~\cite{blumen}; this process leads in
the mean-field limit to a spatially discrete version of the
 well-known Fisher equation~\cite{fisher} 
\begin{equation}
\dot{\phi}_i \ =  r \phi_i (1 - \phi_i) + \frac{D}{a^2} \left(\phi_{i+1} - 2\phi_i
+ \phi_{i-1}\right) . 
\label{fisher}
\end{equation}
Here propagation is into
the linearly unstable $\phi =0$ state and $\phi$ is just the number of
$A$ particles at a site, normalized by $N$. We then assumed that the reaction rate would be a linear function of space, increasing in the direction of propagation,
This type of gradient would be a natural
consequence of spatial inhomogeneity, or could be imposed via a temperature 
gradient in a chemical reaction analog. 
Also, this type of system arises naturally in models of
Darwinian evolution~\cite{tsimring,rouzine}, (where 
fitness $x$ is the independent variable; the birth-rate, akin to
the reaction-rate here, is proportional to fitness). The naive equation describing such a 
model is the Fisher equation (\ref{fisher}) 
with a reaction strength $r=r_a(x)$ varying
linearly in space~\cite{freidlin}
\begin{equation}
r_a(x)  = r_0+ \alpha x \ .
\label{abs}
\end{equation} 
This model
gives rise to an accelerating front. We also introduced a {\em quasi-static} version of the model~\cite{discrete-comment}, 
wherein the reaction rate function moves along with the front:
\begin{equation}
r_q(x)  =  \text{max}(r_{\text min},\tilde{r}_0 + \alpha (x-x_f(t))) \ ,
\end{equation}
with $x_f$ is the instantaneous front position, the precise definition of which we will
discuss later.  The minimum reaction rate $r_{\text min}$ is introduced so as to stabilize the
bulk $\phi=1$ state, and plays no essential role in the following.  This quasi-static problem should lead
 to a translation-invariant front with
fixed speed $v_q (\tilde{r}_0,\alpha )$.   Although important on its own, one might also try to view the
 quasi-static
 problem as a zeroth-order approximation to the original model, (the {\em absolute} gradient case),
  where by ignoring the acceleration, one obtains an adiabatic
approximation to the velocity $v (t; r_0, \alpha ) \simeq v_q
(\tilde{r} _0 (t), \alpha )$ with $\tilde{r} _0 (t) = r_0 +\alpha
x_f (t)$. In both models, the reaction rate gradient was seen~\cite{kess} to
enhance tremendously the role of fluctuations, to the extent that the
naive treatment via a reaction-diffusion, or mean-field, equation gave rise to ``irregular"
behavior completely at odds with the original stochastic model.  In particular,
the reaction-diffusion system exhibited an extreme sensitivity to initial conditions not
present in the stochastic model.  Furthermore, the quasi-static version of the reaction-diffusion
system exhibited a front which accelerated without end, whereas the stochastic version of the
model always achieved an asymptotic constant velocity steady state.  

To get some insight into the stochastic model, we employed a
heuristic approach in which we mimic the leading-order effect of
finite population number fluctuations by introducing a cutoff in the
mean-field equation (MFE)~\cite{mf-dla,kepler,tsimring}. 
This cutoff replaces $r(x)$ by zero if the
density $\phi$ falls below $k/N$ for some $O(1)$ constant $k$; this
change in the reaction term prevents the leading edge from spreading
too far, too fast. This idea has proven its reliability in the Fisher
system with {\em constant} reaction rate where it correctly predicts
the aforementioned anomalous effects~\cite{bd}.  Simulation results~\cite{kess} 
showed that the cutoff MFE does
a quantitatively accurate job of tracking the actual front dynamics.
We then used the cutoff MFE to study the front velocity, at a fixed spatial
position, as
a function of $N$. This was done both for the absolute gradient model
and for the corresponding quasistatic model. From the data, we concluded
that both models exhibit velocities
which increase, evidently without bound, as a function of $N$, 
which is of course
radically different than what had been encountered in the previous
classes of propagating fronts. Thus the cutoff treatment succeeded in
showing why the long-time dynamics of the stochastic model
is not at all correctly described by
the naive reaction-diffusion system.  
In addition, the cutoff theory had the physically reasonable property, again in
accord with the stochastic system, that 
at small enough 
$N$, the velocity could be approximated by just taking a cutoff version
of the usual Fisher equation result for a {\em fixed} reaction rate
$r_F = r_0 +\alpha \bar{x}$, i.e. neglecting the
reaction-rate gradient across the front. This is so because
the effective interfacial width, the distance over which
the particle density drops from its bulk value $O(1)$ to its cutoff
value $O(1/N)$ scales as log $N$; hence one can neglect the gradient
if $\alpha \log{N}$ is small.  The naive reaction-diffusion system, however, due
to its ever increasing interface width, always feels the reaction rate gradient
and never in this adiabatic regime.

Given the highly unusual velocity results, an analytic treatment of
the cutoff system at large $N$ is clearly worthwhile. A very telegraphic
version of this analysis, as applied to the quasi-static model,
was presented in Ref. \onlinecite{kess}.  The
purpose of this paper is to present this analysis in detail.  The continuum
problem is treated first in Sec. \ref{cont}.   A treatment of the dependence of the velocity on the
"base" reaction rate, $r_0$ is presented in the next section. In Sec. \ref{cont-wkb} we redo the continuum
problem via a WKB treatment, developing the methods which will prove necessary
for the lattice problem.  The lattice problem is attacked in Sec. \ref{latt}, producing the controlling
(geometrical optics) WKB approximation, whose properties are then investigated.  The full leading order (physical
optics) WKB solution is obtained in Sec. \ref{wkb2}. This is matched to the solution past the cutoff in Sec. \ref{pastcut},
completing the analysis of the model.  A summary and some concluding remarks then follow.

\section{\label{cont}The Continuum Problem}
 
In this paper we study the steady-state motion of fronts in the quasi-static version of our model.  On the lattice,
the solution has the "Slepyan" traveling wave form ~\cite{slepyan}:
\begin{equation}
\phi_i(t)=\phi(t-ia/v) \ ,
\end{equation}
where the field $\phi_i$ at each lattice site, $i$, has the same history, shifted in time.  The equation of motion then becomes a
differential-difference equation for $\phi$, which we write as a function of the variable $x\equiv -vt$ :
\begin{equation}
\label{full1}
0 = \frac{D}{a^2}(\phi(x+a)+\phi(x-a)- 2\phi(x))   + v\phi'   + r(x)( \phi-\phi^2)\theta(\phi-1/N_e) \ .
\end{equation}
The cutoff in the reaction rate sets is when the density drops below some fraction $k$ of one particle per site, so that $\phi < k/N \equiv 1/N_e$.
We found~\cite{kess} that $k=0.25$ yields excellent quantitative agreement with the stochastic model.
In this quasi-static model, the reaction rate is also only a function of the comoving variable $x$:
\begin{equation}
r(x)=\text{max}(r_{\text min},r_0 + \alpha x) \ ,
\end{equation}
where we have chosen the origin of time such that the front position $x_f$, defined by $\phi_{x_f}=1/2$,
 is located at $i=0$ at $t=0$.   This definition is simplest one for the deterministic problem posed by the cutoff MFE; other conventions are
 more convenient for simulation studies of the stochastic model~\cite{kess}, but this merely corresponds
 to a slight change of $r_0$. 
  
 In the spatial continuum limit, $a \to 0$, this steady-state becomes a standard
 differential equation:  
\begin{equation}
\label{full}
0=D\phi'' + v\phi' + r(x)(\phi-\phi^2)\theta(\phi-1/N_e)
\end{equation}
In this section we treat this simpler problem, returning to the lattice version in Sec. \ref{latt}.

We want to solve the problem for large $N_e$, where as we have
noted we expect the 
velocity to be large.  If $v$ is indeed large, then it appears that the diffusion
term is negligible in comparison, and can be dropped.  We will see that this is in fact valid as long as $x$ is not too large, including the entire
``bulk" region of the solution where $\phi$ is $O(1)$. We then get
\begin{equation}
v\phi' = -r(x)(\phi-\phi^2)\theta(\phi-1/N_e)
\end{equation}
with the solution satisfying $\phi(0)=1/2$ given by
\begin{equation}
\label{bulk}
-\ln\left(\frac{\phi}{1-\phi}\right) = \left\{\begin{array}{cc}
(r_0x + \alpha x^2/2)/v \\
r_{\text{min}} (x-x_{\text{min}})/v + r_0(x_{\text{min}} +
 \alpha x_{\text{min}}^2/2)/v 
 \end{array}\right.
\end{equation}
where the upper term is valid for $ x>x_{\text{min}} $, and the lower term  is
valid for $ x<x_{\text{min}} $.  $x_{\text{min}}$ is the point where the 
minimum reaction rate is reached,
$1+r_0\alpha x_{\text{min}}=r_{\text{min}}$. If we assume for the moment
 that
the solution is valid up to $x_c$, where the cutoff sets in (i.e. $\phi(x_c) =1/N_e$), then all we
have to do is solve for $x>x_c$ and match.  The solution there is
\begin{equation}
\phi = \frac{1}{N_e}e^{-v(x-x_c)/D}
\end{equation}
To do the matching, it is enough to use the
small $\phi$ approximation of Eq. (\ref{bulk}), namely
\begin{equation}
\label{bulk_small}
\phi = e^{-(r_0x+\alpha x^2/2)/v}
\end{equation}.  The matching  of $\phi$ and $\phi'/\phi$ at $x_c$ then gives
\begin{eqnarray}
\frac{1}{N_e}  &=&  e^{-(r_0 x_c+\alpha x_c^2/2)/v} \nonumber \\
\frac{v}{D} &=& \frac{r_0 + \alpha x_c}{v} \ ,
\end{eqnarray}
two equations for the two unknowns $v$ and $x_c$.  For large $N_e$,
both of these are large and we obtain the approximate solution
\begin{eqnarray}
\ln N_e &=&   \alpha x_c^2/(2v) \nonumber \\
v^2/D &=&  \alpha x_c 
\end{eqnarray}
so that
\begin{eqnarray}
v &=& (2D^2 \alpha\ln N_e)^{1/3} \nonumber \\
x_c &=& \left(\frac{4D}{ \alpha}\right)^{1/3}(\ln N_e)^{2/3} 
\label{nobulk_d}
\end{eqnarray}
We can now check our assumption concerning the irrelevance of diffusion
for $x<x_c$.  Using Eq. (\ref{bulk_small}), we find that 
\begin{equation}
\frac{D\phi''}{v\phi'} = D\left[\frac{(r_0 + \alpha x)^2/v^2 + \alpha/v}{(r_0 + \alpha x)}\right]
\end{equation}
For $x$ of order 1, this is of order $1/v$ and is indeed small.  However,
$x_c$ is large, of order $v^2$, so that here the ratio is order 1 and
diffusion can no longer be ignored.  However, since the ratio
is of order 1, and not large, the scaling given by Eq. \ref{nobulk_d} is correct, just not the numerical coefficient.

To incorporate diffusion for $x\lesssim x_c$, we can linearize the
equation since $\phi$ is already small in this region.  We get
\begin{equation}
0=D\phi'' + v\phi' + (r_0+\alpha x)\phi
\end{equation}
Up to a similarity transformation, this is the Airy equation, with
the general solution
\begin{equation}
\phi = e^{-vx/2D}\left[A {\text{Ai}}\left(\frac{\gamma - x}{\delta}\right)
 + B {\text{Bi}}\left(\frac{\gamma - x}{\delta}\right)\right]
\end{equation}
where
\begin{eqnarray}
\gamma &=& \frac{v^2/4D - r_0}{\alpha} \nonumber \\ 
\delta &=& \left(\frac{D}{\alpha}\right)^{1/3}
\end{eqnarray}
We need to match this to the diffusionless solution Eq. (\ref{bulk_small})
for $1 \ll x \ll x_c$, where the arguments of the Airy functions
are large and positive. Doing this, we find that $B$ must be set equal to zero, since 
$\text{Bi}((\gamma-x)/\delta)$ decreases for increasing $x$ and so enhances
the fast descent of the exponential factor.  The $\text{Ai}$ term on the
other hand increases with increasing $x$ and cancels out the fast exponential,
leaving the desired slow exponential of the bulk solution.  Matching to the
bulk solution, we find 
\begin{equation}
A=2\sqrt{\pi}(\gamma/\delta)^{1/4} e^{2/3(\gamma/\delta)^{3/2}}
\label{ai-coef}
\end{equation}
We are now again at a position to perform the match at $x_c$.  The matching
equations are
\begin{eqnarray}
\label{match_full}
\frac{1}{N_e} &=& e^{-vx_c/2D}A {\text{Ai}}\left(\frac{\gamma - x_c}{\delta}\right)
\nonumber \\
\frac{v}{D} &=& \frac{v}{2D} + \frac{\text{Ai}'\left(\frac{\gamma - x_c}{\delta}\right)}
{\delta\text{Ai}\left(\frac{\gamma - x_c}{\delta}\right)}
\end{eqnarray}
Examining the second of this set of equations, we see that the second term
on the left must be large, which we can arrange if the denominator is small;
i.e., $\text{Ai}$ is close to its first zero.
To leading order in $v$, $ \gamma \sim v^2/(4D\alpha) \gg 1$ 
and so $x_c \approx \gamma$. Then, to leading order, we get
\begin{equation}
\ln N_e = \frac{v\gamma}{2D} -\frac{2}{3}\left(\frac{\gamma}{\delta}\right)^{3/2} = \frac{v^3}{24 D^2  \alpha}
\label{lnn-cont}
\end{equation}
so that
\begin{equation}
\label{asym_scaling}
v = (24 D^2  \alpha\ln N_e)^{1/3} 
\end{equation}
so that indeed incorporating the diffusion just modified the prefactor.
Plotting together the exact numerical solution, obtained from a straightforward
Euler initial value integration in time of Eq. (\ref{fisher}), with an 
appropriately small $a$ and $r_q(x)$ as the reaction term,
with the numerical solution to our analytic matching formula Eq. (\ref{match_full}) and our asymptotic scaling solution Eq. (\ref{asym_scaling}), we see
that our matching formula agrees extremely well with the exact velocity. 
Even for the large values of $\ln N_e$ considered here, however, 
the leading order
formula is not very impressive.  The next order term can be calculated and gives
\begin{equation}
\label{asym_corrected}
v = 2(D^2 \alpha)^{1/3}\left[(3\ln N_e)^{1/3}  + \xi_0(3\ln N_e)^{-1/3}\right]
\end{equation}
where $\xi_0=-2.3381$ is the location of the first zero of the Airy function.
Thus, although the correction does decrease with $N_e$, it does so extremely
slowly. This improved approximation is also presented in the figure, 
and does extremely well. 

\begin{figure}
\includegraphics[width=.4\textwidth]{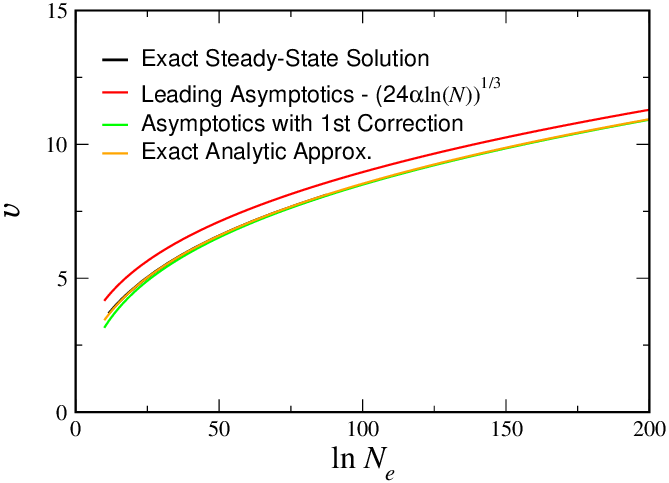}
\caption{Exact quasi-static velocity compared to various approximations for the
spatial continuum case, $a \to 0$, with $\alpha=0.3$, $D=1$, $r_0=1$.}
\end{figure}

\section{Dependence on $r_0$}

\begin{figure}
\includegraphics[width=.4\textwidth]{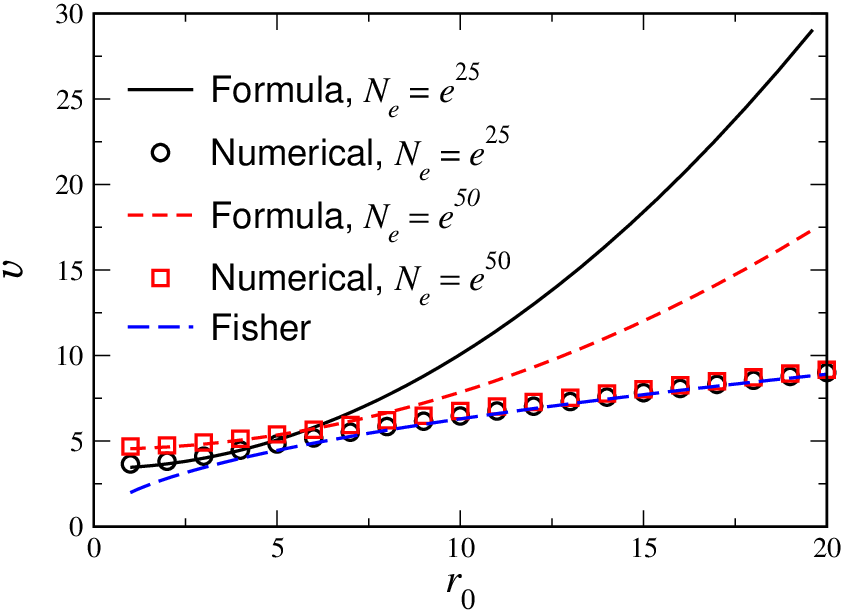}
\caption{Comparison of the predicted $r_0$ dependence, Eq. (\protect{\ref{v_fixed}}), and numerical results for
the spatial continuum limit $a \to 0$, with $ \alpha=0.1$, $D=1$.  Also included is the Fisher velocity, Eq. (\protect{\ref{vfish}}),
where the graphs for the two $N_e$'s coincide on the scale of the figure.}
\label{fig2}
\end{figure}

To the order considered, $r_0$ has not entered into the calculated velocity.  We can calculate the leading $r_0$ dependence by expanding Eq. (\ref{lnn-cont}) to quadratic order in $r_0$:
\begin{equation}
\ln N_e = \frac{v\gamma}{2D} - \frac{2}{3}\left(\frac{\gamma}{\delta}\right)^{3/2} \sim \frac{v^3}{24D^2\alpha} - \frac{r_0^2}{2v\alpha}
\end{equation}
which induces a correction $\Delta v$ to the velocity of 
\begin{equation}
\Delta v =\frac{4D^2 r_0^2}{v^3}
\label{v_fixed}
\end{equation}
so that $v$ increasing quadratically with $r_0$ to leading order.  Note
that the shift is small, of order $1/v$ even for $r_0$ of order $v$.  Also, $r_0$ cannot be taken
to be of order $v^2$, since then diffusion is relevant in the bulk.
In Fig. \ref{fig2} a comparison between formula (\ref{v_fixed}) and numerical results is shown.  The initial quadratic
dependence is clearly seen.  For large enough $r_0$ our formula fails.  Indeed for very large $r_0$, the effect of the
reaction gradient is suppressed, and the velocity should approach that of  the (cutoff) Fisher equation with rate $r_0$~\cite{bd}
\begin{equation}
v = 2\sqrt{r_0D} \left( 1- \frac{\pi^2}{\ln^2{N_e}} \right).
\label{vfish}
\end{equation}

\section{The Continuum Problem, {\it \'a la } WKB}
 \label{cont-wkb}

The fact that we could solve the linear problem exactly obscures the fact
that most
of the structure of the problem comes from the asymptotic properties
of the solution. In fact, we can get essentially everything we require
via a WKB treatment.  Writing $\phi = e^S$, we get to leading order
\begin{equation}
\label{wkb-cont}
DS'^2 + vS' + \alpha y = 0
\end{equation}
where we have written  $y=x + r_0/\alpha$ and
the derivatives are w.r.t. $y$. Equation (\ref{wkb-cont}) defines
$S'$ implicitly in term of $y$.  For small $y$, we get $S'=-\alpha y/v$, so that
$S=S_0 - \alpha y^2/(2v)$.  Matching to the bulk solution gives 
$S_0=r_0^2/(2v\alpha)\approx 0$, if we take $r_0$ to be order 1. Later,
we will see what happens when $r_0$ is of order $v$. Now what is critical, 
as we saw above, is the turning point, since beyond
this point $S'$ turns complex, $\phi$ starts to oscillate, and so hits
zero, which allows us to match to the post-cutoff solution.  The turning
point is given by the discriminant condition, which we can write as
\begin{equation}
\frac{d}{dS'} [DS'^2 + vS' + \alpha y] = 0
\end{equation}
or 
\begin{equation}
S'_* = -\frac{v}{2D}
\end{equation}
Solving for the turning point $y_*$ gives us $y_* = v^2/(4D\alpha)$ which is
consistent with the solution given above in section~\ref{cont}, where the turning point
occurs where the argument of the Ai function is zero.  However, we do not actually
need the value of the turning point, just that of $S'$ there, namely $S'_*$.
Since, as we verify later, the turning point is close to the zero of the solution, 
the dominant contribution to the value of $\phi$ is $e^{S_*}$.
This is given by
\begin{eqnarray}
S_*&=&\int_0^{y_*} dy\,S' = \int_0^{S'_*} dS'\,S' \frac{dy}{dS'} \nonumber\\ 
&= &\int_0^{S'_*} dS'\,S' [-(2DS' + v)/\alpha] = -\frac{v^3}{24D^2\alpha}
\end{eqnarray}
  Note that
we did not need an explicit expression for $S'(y)$, which is good, since
in the lattice case we won't have such an expression.  This calculation
is already enough to give us the leading asymptotics, since
to leading exponential order $\phi(x_c) = e^{S_*}$, or $S_* = -\ln N_e$,
exactly what we got above.  The origin of the correction lies in the
fact that the zero lies a small distance beyond the turning point, namely
$\xi_0 \delta$, a result we need the Airy equation to derive. The real part
of $S'$ is fixed at $S'_*$ beyond the turning point, so $S$ at the zero is
$S_* - \xi_0 \delta v/(2D)$, which we need to set equal to $-\ln N_e$. This
gives us the correction derived above.

\section{The Lattice Problem}
 \label{latt}
 
Now we are in a position to return to our lattice problem, Eq. (\ref{full1}).
As above, in the bulk diffusion is irrelevant and the solution is the same as
before.  Close to the turning point, we linearize and 
expand $S(x\pm a)$ (even though we can't expand $\phi(x\pm a)$), ~\cite{bender,rouzine}
and the WKB equation is
\begin{equation}
0 = \frac{4D}{a^2}\sinh^2(aS'/2) + vS' + \alpha y 
\label{S}
\end{equation}
Already at this point, we get a nontrivial result.  We can de-dimensionalize
this equation by introducing $T = a/\ell S$, $z = y/\ell$, $\ell = v/(a \alpha)$
so that the equation reads
\begin{equation}
\label{scaledwkb}
0 = \frac{4D}{va}\sinh^2(T'/2) + T' + z 
\end{equation}
where the derivative is now w.r.t. $z.$  Thus, $S$ (i.e. $\ln N_e$) scales
like $D/(\alpha a^3)$ times a function of the dimensionless parameter $va/D$,
so that the results for all $a$, (for a given $k$ and $D$) should lie on
a universal curve.  Furthermore, we see that $a$ is
a singular perturbation as far as the large velocity limit goes~\cite{rouzine}, 
since no matter how small $a$ is, the parameter
$va/D$ eventually goes to infinity.

Returning to Eq. (\ref{S}), the turning point is given by the discriminant 
equation
\begin{equation}
0 = \frac{2D}{a} \sinh(aS'_*) + v
\end{equation}
which gives
\begin{equation}
S'_* = \frac{1}{a}\ln\left(\sqrt{1+\frac{v^2 a^2}{4D^2}} - \frac{va}{2D}\right)
\end{equation}
Again, we need to calculate the change in $S$ from $y=0$ to the turning
point $y_*$.  This is given as above by
\begin{eqnarray}
S_* &=& \int_0^{S'_*} dS'\,S' \frac{dy}{dS'}  \nonumber\\
&=& \int_0^{S'_*} dS'\,S' [-(\frac{2D}{a}\sinh(aS') + v)/\alpha] \nonumber\\
&=& -\frac{2D}{\alpha a^2}S'_*\cosh(aS'_*) + \frac{2D}{\alpha a^3}\sinh(aS'_*)
 -\frac{v(S'_*)^2}{2\alpha} 
\label{lat-leading}
\end{eqnarray}
This then is the leading order WKB answer.  Again, to get the correction,
we need to examine the vicinity of the turning point more closely.  We
write
\begin{equation}
\phi(y) = e^{S'_* y} \psi(y)
\end{equation}
In the vicinity of the turning point, this removes the large variation
of $\phi$ between lattice points, leaving us free to Taylor expand the
rest.  Substituting this into Eq. (\ref{full1}), we get
\begin{eqnarray}
0 &=& \frac{D}{a^2}\left[e^{aS'_*}\psi(y+a) + e^{-aS'_*}\psi(y-a) - 2\psi(y)\right] + 
vS'_*\psi(y)  + v\psi'(y) + \alpha y\psi \nonumber \\
&=& \frac{D}{a^2}\left[e^{aS'_*}\left(\psi + a\psi' + a^2 \psi''/2\right)
 +  e^{-aS'_*}\left(\psi - a\psi' +  a^2 \psi''/2\right) 
- 2\psi\right]  +  vS'_*\psi  + v\psi' + \alpha y\psi \nonumber \\
&=& D\cosh(aS'_*)\psi'' + \alpha (y - y_*)\psi
\label{airy_lat}
\end{eqnarray}
Again, we get an Airy equation.  This gives us the distance from the
turning point to the zero of $\psi$, which is 
$\xi_0 \delta_a$ where 
\begin{equation}
\delta_a\equiv(\alpha /(D\cosh(aS'_*)))^{-1/3}
\end{equation}
 is the length scale of the
Airy equation for the lattice problem. This gives us an additional contribution
of $-S'_* \xi_0 \delta_a$ to $S$.  Again, the solution for
the velocity is just $\ln N_e = -S$, so that
\begin{equation}
\ln N_e =\left[\frac{2D}{a^2}S'_*\cosh(aS'_*) - \frac{2D}{a^3}\sinh(aS'_*) 
+ v(S'_*)^2/2\right]/\alpha 
 -  \xi_0 S'_* \left(\frac{\alpha}{D\cosh(aS'_*)}\right)^{-1/3}
\label{lat-correct}
\end{equation}

Let us examine the various limits of the this expression.  First, the
continuum limit, $av/D \ll 1$.  Then $S'_*=-v/2D$, as in
the continuum calculation, so that $aS' \ll 1$.
Then, $S_* = -(2D(S'_*)^3/3 + v(S'_*)^2/2)/\alpha =-v^3/(24D^2\alpha)$, also exactly as in the continuum
calculation.  The correction term is $-S'_* \xi_0 (\alpha/D)^{-1/3}$, which
also agrees.

\begin{figure}
\includegraphics[width=.4\textwidth]{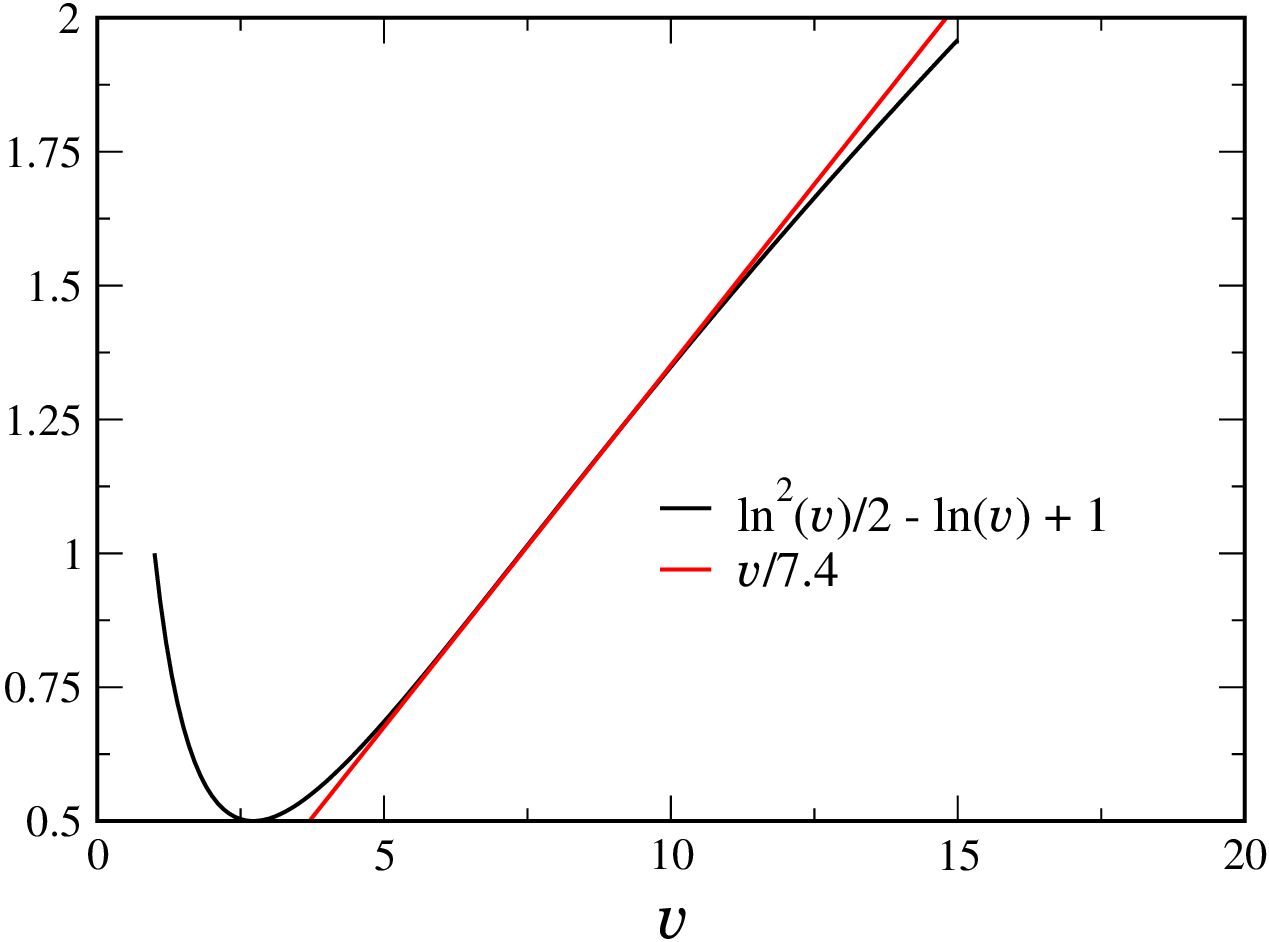}
\caption{Comparison of the function $\ln^2(v)/2 - \ln(v) +1$ and the 
linear approximation $v/7.4$}
\label{linear}
\end{figure}

Now, as we mentioned above, for any finite $a$, $av/D$ is eventually large
for sufficiently large $N_e$. Then, $S'_* = -\ln(va/D)/a$.  This gives
$S_* = v/(a^2 \alpha) (-\ln(va/D) + 1 + \ln^2(va/D)/2)$.  
Now, for very large $va/D$, $S_* \approx -v\ln(va/D)/(a^2 \alpha)$.  However,
this is only valid for $ln(va/D) \gg 2$. In fact, it is a reasonable (20\%)
approximation only for $\ln(va/D)$ bigger than 10, so that $v$ would be 
unreasonably large.  Thus, a strict asymptotic expansion is of no use
whatsoever.  Over the range $5 < x < 11$, an excellent approximation
of $\ln^2(x)/2 + 1 - \ln(x)$ is $x/7.4$ (see Fig. \ref{linear}).  

Thus, in the relevant range,
$S_* \approx v^2/(7.4 a \alpha D)$. Thus, while formally, 
to leading order 
\begin{equation}
\label{leading-leading}
\ln N_e \approx v/(a^2 \alpha) \ln^2(va/D)
\end{equation}
or equivalently
\begin{equation}
v \approx a^2 \alpha \ln(N_e)/ \ln^2(a^3 \alpha\ln(N_e)/D)
\end{equation}
this is true only for astronomically large $N_e$.  More useful,
though phenomenological, is
\begin{equation}
\label{simple}
v \approx \sqrt{7.4 a \alpha D\ln N_e}
\end{equation}
Thus, the velocity increases with $\alpha$, $D$ and $N_e$. Including the
correction term, we get the effective approximation
\begin{equation}
\label{very_approx}
v \approx \sqrt{7.4 a\alpha D\ln N_e}
 - 1.76 \alpha ^{1/3} D^{2/3}(\ln(N_e))^{-1/3}\ln(7.4 a^3\alpha \ln(N_e)/D)
\end{equation}

\begin{figure}
\includegraphics[width=.4\textwidth]{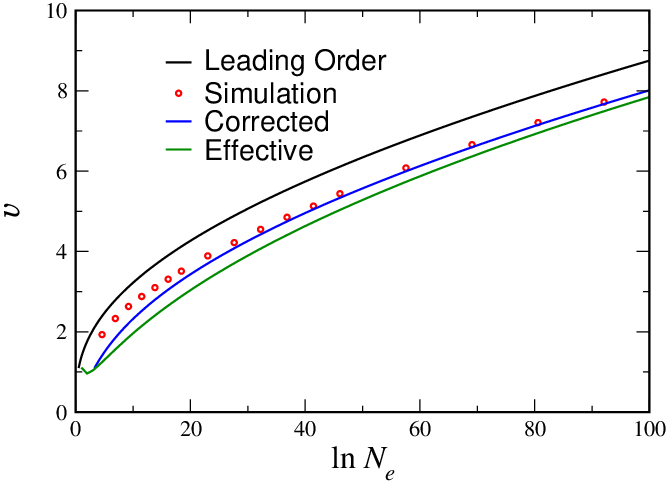}
\caption{Velocity vs. $\ln(N_e)$ for lattice spacing $a=1$, $\alpha=0.1$, $D=r_0=1$ from simulation, together with the various analytic approximations.
 The curve labeled ``Leading Order'' represents Eq. (\protect{\ref{lat-leading}}) and that labeled ``Correction'' represents
  Eq. (\protect{\ref{lat-correct}}). The curve labeled ``Effective'' represents Eq. (\protect{\ref{very_approx}}).}
\label{fig4ac}
\end{figure}

\begin{figure}
\includegraphics[width=.4\textwidth]{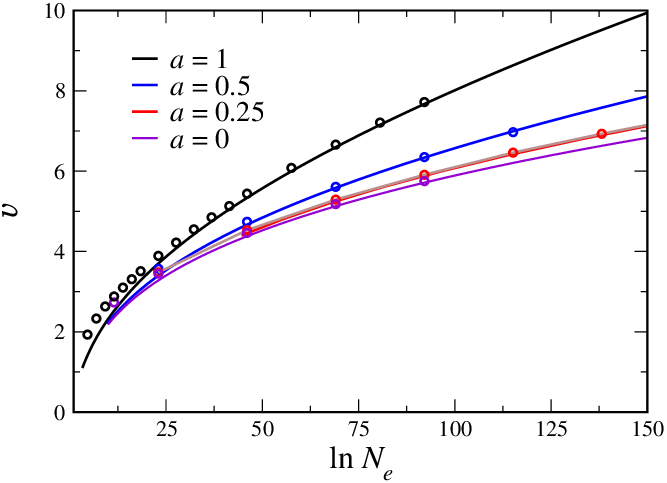}
\caption{Velocity vs. $\ln(N_e)$ for lattice spacings $a=1$, 0.5, 0.25, and 0,
$\alpha=0.1$, $D=r_0=1$ from simulation, together with the analytic approximation Eq. (\protect{\ref{lat-correct}})}
\label{fig5ac}
\end{figure}

 Fig. \ref{fig4ac} presents the case $a=1$.
We see that for $v$'s bigger than 4, the corrected approximation
is excellent.  The leading order approximation, however, is poor
 even for $\ln(N_e)$ as unreasonably large as 100.  The
simplified effective approximation Eq. (\ref{very_approx}) is as
good as the full corrected approximation for this range of $N_e$.  The
extremely simple Eq. (\ref{simple}) is as good as the leading order
approximation.
In Fig.
\ref{fig5ac}, we present data for a number of values of $a$, ranging
from 0 to 1, along with our analytic prediction.  Again, the agreement
is very good.

\begin{figure}
\includegraphics[width=.4\textwidth]{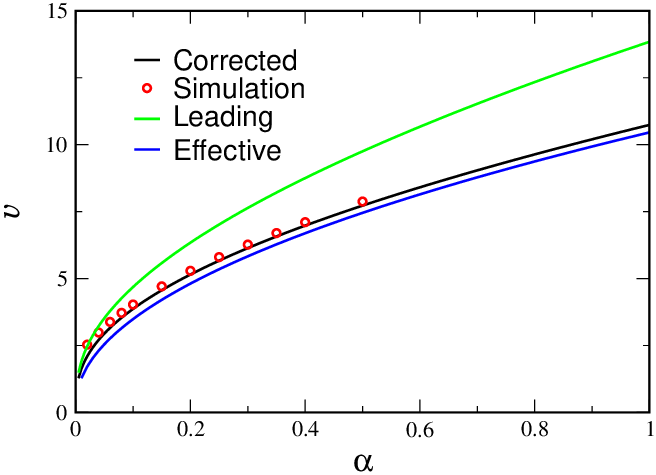}
\caption{Velocity vs. $\alpha$ for lattice spacing $a=1$,
$D=r_0=1$, $\ln(N_e)=25$, from simulation, together with the leading-order
and corrected analytic approximations, as in Fig. \protect{\ref{fig4ac}}.}
\label{fig_many_alf}
\end{figure}

In Fig. \ref{fig_many_alf}, we show the dependence on $\alpha$, 
the gradient of the reaction rate.  We see that the rise in $v$ is quite
steep at first, and then tapers off to an much slower rise.  It should
be noted how much an effect the correction term has, especially at larger
$\alpha$. Nowhere does it look simply proportional to $\alpha$, as would
naively appear from the leading order calculation, Eq. (\ref{leading-leading}).

\begin{figure}
\includegraphics[width=.4\textwidth]{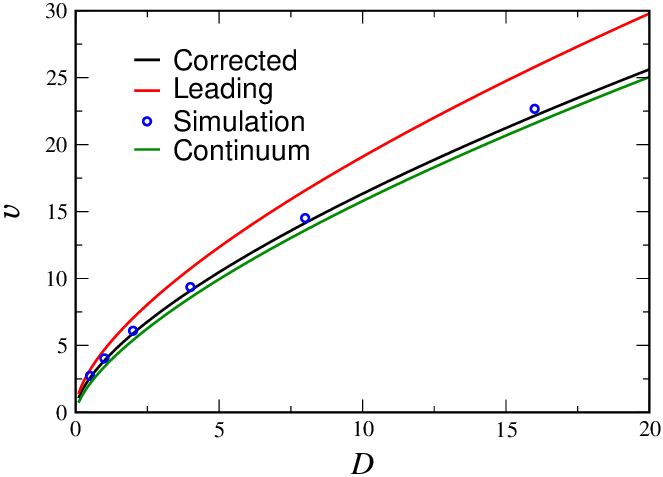}
\caption{Velocity vs. $D$ for lattice spacing $a=1$,
$\alpha=0.1$, $r_0=1$, $\ln(N_e)=25$, from simulation, 
together with the leading-order
and corrected analytic approximations, as in Fig. \protect{\ref{fig4ac}}.
Also included is the continuum approximation, Eq. (\protect{\ref{asym_corrected}}).}
\label{fig7ac}
\end{figure}

In Fig. \ref{fig7ac}, we show the dependence of $v$ on the diffusion 
constant, $D$.  Again, the rise in $v$ is steep at small $D$, 
and grows essentially linear for large $D$.  As $v/D$ is a decreasing
function of $D$, for large $D$ the continuum limit is eventually
valid.  We see in fact that the continuum approximation, Eq. (\ref{asym_corrected}) works quite well over the entire range of $D$.

\section{Next-order WKB}
 \label{wkb2}

To go further, we need to both improve our WKB solution and to extend
our solution to the region $x>x_c$.  Note that it is easy to check that we do not need to reconsider the diffusionless solution in the bulk, as the first-order correction to that solution  is lower order (in terms of its ultimate effect on the velocity) than the correction we derive here. We first consider the next-order
WKB solution.  We write $\phi = e^{S_0 + S_1}$, and get the next order
WKB equation
\begin{equation}
\frac{2D}{a^2}\left[aS_1' \sinh(aS_0') + \frac{a^2}{2}S_0'' \cosh(aS_0')
\right] + vS_1' = 0
\end{equation}
with the solution
\begin{equation}
S_1 =  -\frac{1}{2}\ln\left(\frac{av}{2D} + \sinh(aS_0')\right) + C_1
\end{equation}
so that
\begin{equation}
\label{sol_e}
\phi \approx  (av/2D + \sinh(aS_0'))^{-1/2} e^{C_1 + S_0}
\end{equation}
We have to match the solution to the bulk solution, which is approximately 
$e^{-\frac{\alpha y^2}{2v}}$. The matching area is defined by the requirement
that on the one hand diffusion be irrelevant, so that $-S_0' \approx
\alpha y/v \ll \ln v$, or $y \ll v\ln v$, 
and on the other, $\phi$ is small, implying $y^2/v \gg 1$, or $y \gg v^{1/2}$.
Thus to do the matching we can take $y \sim  O(v)$.
Neglecting the $\sinh$, we get from Eq. (\ref{sol_e}):
\begin{equation}
C_1=\frac{1}{2}\ln\left(\frac{va}{2D}\right),
\label{sol_C}
\end{equation}
and the WKB solution is,
\begin{eqnarray}
\phi=e^{S_0}\left(1 + \frac{2D}{va}\sinh(aS')\right)^{-1/2}
\label{sol_WKB}
\end{eqnarray}

We now need to match the WKB solution to the Airy solution of $\phi=e^{S_*'(y-y_*)}\psi$, where $\psi=C_2 Ai((y_*-y)/\delta_a)$.
 First we find the matching region.
Clearly we need the argument of the Airy function to be large.  Thus gives us
that $y_*-y \gg \delta_a$.  In the lattice limit, this reduces to
$y_*-y \gg v^{-1/3}$, while in the continuum limit, we get
$y_*-y \gg 1$. Near the turning point, 
$S_0' \approx S_*' + \delta_a^{-3/2}\sqrt{y_*-y}$. This is valid as long as
$a\delta_a^{-3/2}\sqrt{y_*-y} \ll 1$, or $y_* - y \ll \delta_a^3/a^2$.  Thus,
we can match as long as $\delta_a \ll \delta_a^3/a^2$, which is
uniformly true in the large $v$ limit. 
Approximating the Airy solution, we get
\begin{eqnarray}
\phi&\approx &C_2 e^{S_*'(y-y_*)}\textrm{Ai}\left((y_*-y)/\delta_a\right)
\nonumber\\ 
&\approx& 
\frac{1}{2\sqrt{\pi}}\delta_a^{1/4}(y_*-y)^{-1/4}C_2 e^{S_*'(y-y_*)}
e^{-\frac{2}{3}\left((y_*-y)/\delta_a\right)^{3/2}}
\end{eqnarray}
Approximating the WKB solution, we get
\begin{eqnarray}
\phi&\approx&\left(1+\frac{2D}{va}\sinh(aS_0'(y))\right)^{-1/2}\nonumber e^{S_0(y)}\\
&\approx& \left(1 + \frac{2D}{va}\left[\sinh(aS_*') + a\cosh(aS_*')\sqrt{\delta_a^{-3}(y_*-y)}\right]\right)^{-1/2} 
 e^{S_* - S_*'(y_*-y) - \frac{2}{3}\delta_a^{-3/2}(y_*-y)^{3/2}} \nonumber\\
&\approx& \left(\frac{2D}{v}\cosh(aS_*')\sqrt{\delta_a^{-3}(y_*-y)}\right)^{-1/2}
 e^{S_* - S_*'(y_*-y) - \frac{2}{3}\delta_a^{-3/2}(y_*-y)^{3/2}} 
\label{matching}
\end{eqnarray}
Matching these two gives
\begin{equation}
C_2=\frac{\sqrt{2\pi v}}{D^{1/3}\alpha^{1/6}\cosh^{1/3}(aS_*')}e^{S_*}
\label{C2}
\end{equation}
It is easy to verify that this agrees in the $a \rightarrow 0$ limit with the direct continuum calculation,
Eq. (\ref{ai-coef}).

\section{Matching to $x>x_c$}
 \label{pastcut}
Finally, we have to actually match to the
solution for $x>x_c$.  For $x>x_c$, we clearly cannot expand $\psi$ as
we have done, since the falloff of $\phi$ is much faster than $exp(-S'_*x)$.
We can understand what happens by considering doing the expansion of $\psi$
to one more order.  The higher derivative induces a boundary layer at $x_c$,
which serves to insure continuity of the second derivative, but leaves
the lower-order derivatives untouched.  As we expand to higher and higher
order, there are more and more boundary-layer modes, which we have to
match.  The correct way to do the matching then is via a Wiener-Hopf (WH)
procedure. To do the WH, we consider our problem in the immediate vicinity of
$x_c$.  Here, we can approximate $\alpha y$ by $\alpha y_c$, since as we have seen,
$y_c$ is large.  Now we have a constant coefficient difference equation,
which we can solve via WH. 

Let us do this first for the continuum problem for practice, since
here we know the correct answer.  The approximate equation is
\begin{equation}
D\phi'' + v\phi' + \alpha y_c\theta(y_c-y)\phi = 0
\end{equation}
Writing $\phi=\phi_L + \phi_R$ and Fourier transforming, we get
\begin{equation}
-Dq^2(\phi_L + \phi_R) + ivq(\phi_L + \phi_R) + \alpha y_c\phi_L = 0
\end{equation}
or equivalently
\begin{equation}
(-Dq^2 + ivq)\phi_R = -(-Dq^2 + ivq + \alpha y_c)\phi_L 
\end{equation}
The right-hand operator $R(q)=-Dq^2+ivq$ has two zeros, one at $q=0$
and one at $q=iv/D$, but  $\phi_R$ has a pole only at $q=q_R=iv/D$. The
left-hand operator $L(q)=-Dq^2 + ivq + \alpha y_c$ has two essentially
degenerate zeros, close to $q=iv/(2D)$, both of which are 
represented as poles in $\phi_L$. Thus, we rewrite the equation as follows
\begin{equation}
(q - q_R) \phi_R = -\frac{L(q)(q-q_R)}{R(q)}\phi_L
\end{equation}
Now, the left-hand side of the equation has no zeros or poles below
the line $\Im{q}=v/D$, and the right-hand side of the equation
has no zero or poles above, so they must both be equal to a constant, $C$.
Thus,
\begin{eqnarray}
\phi_R &=& \frac{C}{q - q_R} \nonumber \\
\phi_L &=& -\frac{C R(q)}{L(q)(q-q_R)} = -\frac{Cq}{(q-q_+)(q-q_-)}
\end{eqnarray}
where $q_\pm$ are the two nearly degenerate roots of $L(q)$, 
\begin{equation}
q_{\pm} = \frac{iv}{2D} \pm \Delta
\end{equation}
with $\Delta=\frac{\sqrt{-v^2 + 4D\alpha y_c}}{2D}$ small. 
Fourier transforming back, we get
\begin{eqnarray}
\phi_R &=& iC e^{-v(y-y_c)/D} \nonumber \\
\phi_L &=& iC\left[\frac{q_+}{q_+ - q_-} e^{iq_+ (y-y_c)}
+ \frac{q_-}{q_- - q_+} e^{iq_- (y-y_c)}\right]
\end{eqnarray}

Examining $\phi_R$, we find that $iC=1/N_e$.  Turning to $\phi_L$,
we found above that $y_c \approx v^2/(4\alpha D) - \xi_0 \delta$, so
\begin{equation}
\Delta \approx \sqrt{-\alpha  \xi_0 \delta/D}
\end{equation}
Now, as long as $(y-y_c)\Delta \ll 1$, we can write
\begin{equation}
\phi_L = \frac{ e^{-v(y-y_c)/(2D)}}{N_e} \left[ 1 - \frac{v}{2D}(y-y_c) \right]
\end{equation}
We now have to match this to the Airy function.  Putting $y_c = v^2/(4D\alpha) - 
\xi_0\delta - y_1$, we find that
\begin{eqnarray}
\phi &\approx& A e^{-vy_c/2D}e^{-v(y-y_c)/(2D)} \text{Ai}'(\xi_0)\left(\frac{y_1 - (y-y_c)}{\delta}\right) \nonumber \\
&=& A e^{-vy_c/2D} e^{-v(y-y_c)/2D}\frac{\text{Ai}'(\xi_0) y_1}{\delta} \left(1 - \frac{y-y_c}{y_1}\right) 
\end{eqnarray}
Comparing the two results we find $y_1=2D/v$, and
\begin{equation}
A e^{-vy_c/2D} \frac{2D\text{Ai}'(\xi_0)}{v\delta} = \frac{1}{N_e}
\end{equation}
or
\begin{eqnarray}
\ln N_e&=& -\frac{2}{3}\frac{v^3}{(4D\alpha )^{3/2}}\left(\frac{D}{\alpha}\right)^{-1/2} +  \frac{v^3}{8D^2\alpha}  - \frac{v\xi_0\delta}{2D}
 - 1 -\frac{1}{2}\ln 4\pi - \frac{1}{4}\ln\left(\frac{\gamma}{\delta}\right) - \ln\left(\frac{2D\text{Ai}'(\xi_0)}{v\delta}\right) \nonumber \\
&=&\frac{v^3}{24 D^2 \alpha} - \frac{v\xi_0}{2 (D^2 \alpha)^{1/3}} - \ln\left(\text{Ai}'(\xi_0) 2e \sqrt{2\pi} D^{1/3} \alpha^{1/6}  v^{-1/2} \right)
\end{eqnarray}
This can be shown to agree with the direct asymptotic solution of Eq. \ref{match_full}.

\begin{figure}
\includegraphics[width=.4\textwidth]{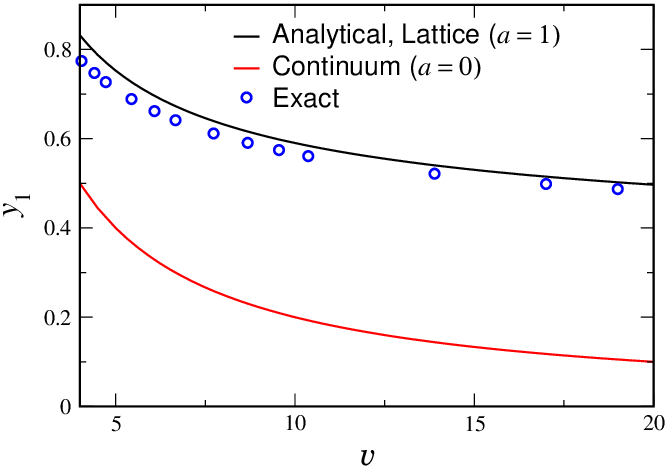}
\caption{Large $v$ analytic approximations for $y_1$ vs. velocity for 
$\alpha=0.1$, $r_0=1$, $D=1$, both for lattice spacing $a=1$,
and for the continuum limit, together with results from simulation.}
\label{y1fig}
\end{figure}

Now we do the same for the lattice problem.  Again, after
setting $\alpha y=\alpha y_c$, the equation has the form
\begin{equation}
R(q)\phi_R = - L(q)\phi_L
\end{equation}
where
\begin{eqnarray}
R(q) &=& -\frac{4D}{a^2}\sin^2\frac{aq}{2} + ivq \nonumber \\
L(q) &=& R(q) + \alpha y_c
\end{eqnarray}
$R(q)$ has two pure imaginary roots, one at $q=0$ and the other with
positive imaginary part, $\kappa_R$.  
Since we cannot allow $\phi_R$ to become
negative, the dominant solution for large $x$ must be controlled
by this positive imaginary root, leading to a pure exponential decay.
Only this root and the complex roots which decay faster ($\Im q > \kappa_R$)
are then permissible.
As in the continuum, since $y_c$ is close to $y_*$, $L(q)$ has a pair
of almost degenerate roots, $q_\pm$ with small real parts and a 
positive imaginary
part smaller than $\kappa_R$.  To match to the $Ai$ solution, this must be the dominant contribution
for large negative $y-y_c$, so the only acceptable roots
are those with $\Im q \le \Im q_\pm$.
We therefore decompose $R(q)$ and $L(q)$ into two factors, one with
its zeros below $i\kappa_R$, which we label by a ``$B$'' superscript,
and the other with its zeros above (or equal), which we label by ``$U$''.
Formally,
\begin{eqnarray}
R(q)&=&R^U(q)R^B(q)\nonumber\\
L(q)&=&L^U(q)L^B(q)
\end{eqnarray}
where
\begin{eqnarray}
R^U(q)&=& \prod_i \left(1-\frac{q}{q^U_{R,i}}\right) \nonumber\\
R^B(q)&=& ivq\prod_i \left(1-\frac{q}{q^B_{R,i}}\right) \nonumber\\
L^U(q)&=& \alpha y_c\prod_i \left(1 - \frac{q}{q^U_{L,i}}\right)\nonumber \\
L^B(q)&=& \left(1-\frac{q}{q_+}\right)\left(1-\frac{q}{q_-}\right)
\prod_i \left(1 + i\frac{q}{q^B_{L,i}}\right)
\end{eqnarray}
We have chosen to explicitly break out the factors relating to $q_\pm$ in 
$L^B$ since they will play an essential role in the following, and the
factor $ivq$ in $R^B$ so that the correct behavior at $q=0$ is maintained.
Then, via the standard Wiener-Hopf argument,
\begin{eqnarray}
\phi_R(q) &=& C\frac{L^U}{R^U} \nonumber \\
\phi_L(q) &=& -C\frac{R^B}{L^B}
\end{eqnarray}
It is easiest to proceed if we regularize the problem by effectively
discretizing time, replacing the $ivq$ term in the operators $L$, $R$
by $iv n_t/a \sin(qa/n_t)$ where $n_t$ is some large integer.  Then, the
operators become polynomials of order $2n_t$ in the variable $e^{iqa/n_t}$.
Further study reveals that $R^U$ then has $n_t-1$ zeros, $R^B$ has $n_t+1$,
$L^U$ has $n_t-2$ zeros, and $L^B$ has $n_t+2$. Thus, $\phi_L$ behaves
as $q^{-1}$ for large $q$, and so 
\begin{equation}
\phi_L(y_c) = -i\lim_{q\to \infty} q\phi_L(q) =
 -Cvq_+q_-\prod \frac{q^B_{L,i}}{q^B_{R,i}}
\end{equation}
so that
\begin{equation}
C = -\frac{1}{N_{e}vq_+q_-}\prod \frac{q^B_{R,i}}{q^B_{L,i}}
\end{equation}
so that
\begin{equation}
\phi_L(q) = \frac{iq}{N_e(q-q_+)(q-q_-)} \prod_i \frac{q-q^B_{R,i}}{q-q^B_{L,i}}
\end{equation}
If $y$ is not too close to $y_c$, only the two dominant modes, which we
have labeled $q_{\pm}=-iS'_* \pm \Delta$ survive,
and we get
\begin{eqnarray}
\phi_L(y) \approx \frac{1}{N_e}\prod_i \frac{-iS'_* - q^B_{R,i}}
{-iS'_* - q^B_{L,i}} e^{S'_*(y-y_c)}\left(1
+ \sum_{j} \left(\frac{-S'_*}{-S'_* + i q^B_{R,j}} 
                  - \frac{-S'_*}{-S'_* + i q^B_{L,j}}\right)
+ S'_*(y-y_c) \right)
\label{wh}
\end{eqnarray}
This is seen to reproduce the continuum results above when $a\to 0$.  
Now we must match Eq. (\ref{wh}) to
our Airy function solution to Eq. (\ref{airy_lat}).  Actually, to the order we are working, we
must take into consideration the first lattice correction to the Airy equation, namely
\begin{equation}
\frac{Da\sinh(aS'_*)}{3}\psi''' + D\cosh(aS'_*) \psi'' + \alpha(y-y_*)\psi = 0 \ .
\end{equation}
The (first order) approximate solution to this is
\begin{equation}
\psi(y) \approx C_2 \left(1+\frac{a\tanh(aS'_*)}{12\delta_a^3}(y-y_*)^2\right) \text{Ai}\left(-\frac{y-y_*}{\delta_a} + 
\frac{a\tanh(aS'_*)}{6\delta_a}\right) \ . 
\end{equation}
Substituting $y_c = y_* - \xi_0 \delta_a - y_1$, with $y_1/\delta_a \ll 1$, we find
\begin{equation}
\phi(y) \approx C_2 e^{S'_*(y_c-y_*)}e^{S'_*(y-y_c)} \frac{Ai'(\xi_0)}{\delta_a} 
\left[ y_1 + \frac{a\tanh(aS'_*)}6 - (y-y_c) \right] \label{final-airy}
\end{equation}
Thus gives us
\begin{equation}
y_1 = -\frac{a\tanh(aS'_*)}{6} + \frac{1}{-S'_*} + \sum_j \left( \frac{1}{-S'_* + iq^B_{R,j}}
- \frac{1}{-S'_* + iq^B_{L,j}} \right)
\end{equation}
A graph of $y_1$ as a function of $v$ is presented in Fig. \ref{y1fig}, 
together
with the continuum result. We see that whereas $y_1$ falls with $v$ in the
continuum, due to the lattice correction to the Airy equation, the lattice $y_1$ approaches
a constant for large $v$.  In fact, at large $v$, all the $q$'s can be
calculated analytically, and the some performed.  This calculation shows
that to leading order, both the sum over the right and left modes approaches
$1/2$, with the difference vanishing as $v \to \infty$. The lattice $y_1$
is dominated by the lattice correction to the Airy equation, which
approaches the constant $a/6$ for large 
$v$.
Included in this figure is a comparison between analytical and numerical results.  We see that as
expected the analytic result approaches the numerical results are $v$ increases, being quite
accurate everywhere.

The numerical results presented in this graph, in contrast to those presented throughout the rest of this paper,
 were not obtained through direct numerical
simulation of the time-dependent equations, due to the high accuracy required to perform the comparison with the
theory.  Our direct numerical simulations were performed via a straightforward Euler simulation, which is only
first-order accurate in the time step.  Extremely small time-steps would have been required to obtain the
requisite accuracy.  Instead, we solved the linearized steady-state equation directly.  As opposed to $v(N_e)$, which
requires a full nonlinear solution, $y_1(v)$ is determined solely by the linearized equation.  The procedure we employed
was as follows:  The solution past the cutoff was written as a linear superposition of the allowed modes, corresponding to the
roots $q_R^U$.  The solution to the left of some conveniently chosen $y_\ell$ was written as a linear superposition of modes, with
the reaction rate set at the constant value $r(y_\ell)=r(y_c)-\alpha(y_c-y_\ell)$.  The steady-state equations between $y_\ell$ and $y_c$
were written as a banded matrix, acting on the three sets of unknowns: the coefficients of the pre-$y_\ell$ modes; the values of the
field between $y_\ell$ and $y_c$; and the coefficients of the post-$y_c$ modes.  This matrix depends on the two parameters $v$ and
$r(y_c)$.  Now the pre-$y_\ell$ modes include two real modes, corresponding to the two solutions of the quadratic equation
for $S'$.  In order to match the Airy function behavior, the faster of these
two modes must not be present.  This is an eigenvalue condition of $r(y_c)$ for a given $v$.  From $r(y_c)$ we can back out $y_1$, as
presented in the figure.  This procedure converges quadratically in the discretization $\Delta y$.  The convergence with respect to
$y_\ell$ is exponentially rapid, and presented no problem. A comparison of the answers obtained in this manner with that obtained
by direct numerical simulation, at low values of $v$ for which the latter calculation was feasible, verified the validity of this alternate
approach.  This new approach also sheds an interesting light on the selection problem inherent in the linearized steady-state equation.

\begin{figure}
\includegraphics[width=0.4\textwidth]{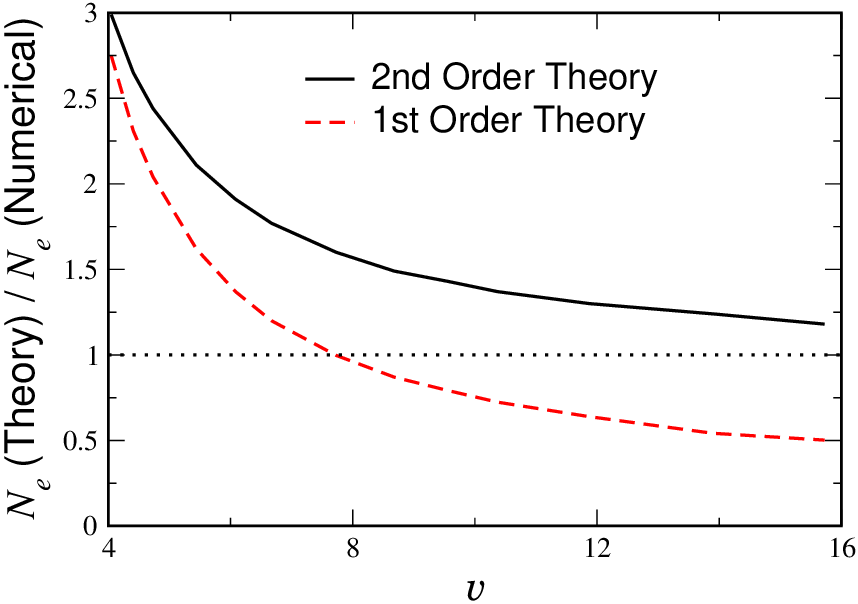}
\caption{Ratio of $N_e$ as predicted by Eq. (\protect{\ref{final}}) to the exact $N_e$ from numerical
simulation for $D=1$, $\alpha=0.1$, $a=1$.  Also shown is the ratio of $N_e$ as predicted from the
lower-order result, Eq. (\protect{\ref{lat-correct}}) to the exact $N_e$.}
\label{nratio}
\end{figure}

All that remains is to put everything together and construct the full approximation for the
velocity as a function of the cutoff, $N_e$.  
As we have seen, in the matching region, the WH $\phi_L$ has the
same functional form as the Airy solution, once $y_1$ is picked appropriately
as described above, the two solutions differing only in normalization.
Setting the normalization factors equal then fixes $1/N_e$ in terms of
our previously calculated $C_2$, (see Eq. (\ref{C2})). Doing this gives
\begin{equation}
\ln N_e = -S_* + S'_*(\xi_0 \delta_a + y_1) - \ln\left(\sqrt{2\pi v/\alpha}{\delta_a}^{-1}\right)
 - \ln\left(\frac{{\text{Ai}}'(\xi_0)}{\delta_a (-S'_*)} \prod \frac{-iS'_* - q^B_{L,i}}{-iS'_* - q^B_{R,i}}\right)
\ .
\label{final}
\end{equation}
In Fig. \ref{nratio} we present the ratio of $N_e$ as predicted by this formula to the results
of numerical simulation.  We see that the ratio appears to approach unity as $v$ increases.
Together with this is shown the ratio of $N_e$ from the lower order formula, Eq. (\ref{lat-correct}).
This formula, while it does better at small $v$, is seen to diverge from the exact answer with
increasing $v$.

\section{Conclusions}
In summary, we have presented an analytical study of the velocity of Fisher 
fronts in the presence of a gradient.  This study  exploits the fact that the
 velocity diverges as the local density of reactants increases.  This 
divergence is one of the signposts of the extreme sensitivity to fluctuations 
of this class of models. One of the most surprising consequences of this 
sensitivity is the different order of the divergence in the continuum versus 
the lattice model; whereas the velocity of the front in the continuum limit 
diverges as $\ln^{1/3}(N)$,on the lattice, the velocity effectively diverges as
$\sqrt{\ln{N}}$. The relative insensitivity to the details of the matching to 
the post-cutoff regime is another characteristic feature of
this problem.  It is clear, for example, that the leading order results are 
completely independent of the post-cutoff dynamics. Even the next order, which 
formally does depend on matching to the solution past $x_c$,  is in fact only 
very weakly modified by the ``R" modes arising from this region. This lack of 
strong dependence is no doubt a major part of the reason that the 
phenomenological cutoff theory works
as well as it does in describing the stochastic model.

Obviously, the cutoff MFE approach cannot capture any of the truly stochastic 
features of the original Markov model. Thus, the next step in our overall 
program for understanding fronts in gradients must involve adding back in the 
residual effects of finite particle number fluctuations to the cutoff theory. 
Exactly how to do this is already unclear in the simpler case of the Fisher 
equation front, where it has proven difficult to come up with a simple 
explanation for the numerically determined front diffusion 
constant.~\cite{bd2,debrata}. The first question to be answered for the 
gradient case is whether the front can be described as simply diffusing 
(albeit with an anomalous diffusion constant) or whether the fluctuation 
effects perhaps lead to even stronger stochasticity. We hope to report on this 
issue in a future publication.

\acknowledgments{EC and DAK acknowledge the support of the Israel Science 
Foundation.  The work of HL has been supported in part by the NSF-sponsored 
Center for Theoretical Biological Physics (grant numbers PHY-0216576 and 
PHY-0225630).}

\end{document}